\documentclass[12pt]{article}
\pdfoutput=1
\usepackage[top = 2 cm, bottom = 3 cm, left = 2.5 cm, right = 1.5 cm]{geometry}
\usepackage{authblk, cite, color, amssymb, amsmath, graphicx}
\usepackage[colorlinks = true, linkcolor = blue, citecolor = red]{hyperref}
\usepackage[dvipsnames]{xcolor}

\begin{document}

\title{Supplying Dark Energy from Scalar Field Dark Matter}

\author{{\bf Merab Gogberashvili$^{1,2}$} and {\bf Alexander Sakharov}$^{3,4,5}$}
\affil{\small $^1$ Javakhishvili Tbilisi State University, 3 Chavchavadze Avenue, Tbilisi 0179, Georgia \authorcr
$^2$ Andronikashvili Institute of Physics, 6 Tamarashvili Street, Tbilisi 0177, Georgia \authorcr
$^3$Department of Physics, New York University, New York, NY 10003, USA \authorcr
$^4$Physics Department, Manhattan College, Riverdale, New York, NY 10471, USA \authorcr
$^5$EP Department, CERN, 1211 Geneva 23, Switzerland}
\maketitle

\begin{abstract}

We consider the hypothesis that dark matter and dark energy consists of ultra-light self-interacting scalar particles. It is found that the Klein-Gordon equation with only two free parameters (mass and self-coupling) on a Schwarzschild background, at the galactic length-scales has the solution which corresponds to Bose-Einstein condensate, behaving as dark matter, while the constant solution at supra-galactic scales can explain dark energy.

\end{abstract}
\vskip 5mm


Although, the $\Lambda$CDM model explains very successfully numerous observations on both the large scale structure and galaxies formation, the microscopic nature of the Dark Matter (DM) is still unknown. In the most common paradigm, DM constituents are so called weakly interacting massive particles. However, attempts to detect those yet allowed only obtaining restrictions on the basis of constraints coming from non-observations of signals over backgrounds of different instruments. Moreover, it turns out that although the standard $\Lambda$CDM can successfully accommodate the large scale structure formation through the evolution of DM fluctuations, there are still some poorly understood issues on galactic and sub-galactic scales, such as cuspy halos~\cite{cdmProb1, cdmProb2, coreHalo1}, missing satellite problem~\cite{cdmProb3} and too big to fail problems~\cite{cdmProb4, cdmProb5, cdmProb6}.

All mentioned small-scale structure anomalies can in principle be resolved if the DM is made up of ultra-light bosons that form a Bose-Einstein Condensate (BEC), i.e. a single coherent macroscopic wave function with long range correlation~\cite{Lee_first,BBM, SRM, shapiro1, shapiro2, shapiro3, MGN, FM, DB, CEMN, BH, DKG}. Such kinds of bosons are usually modeled by a scalar field either with or without self-interaction. In case the scalar field DM does not possess any self-interaction, the quantum pressure of localized particles would be sufficient to stabilized the DM halo against gravitational collapse only for very light particles with mass $m\sim 10^{-22}~{\rm eV}$~\cite{noSelf1, noSelf2, noSelf3, noSelf4, noSelf5}. From the other side, introduction of a small repulsive self-interaction makes possible to extend the range of allowed masses of the scalar DM particles up to $m\le 1~{\rm eV}$~\cite{self1, Goodman1, Julian1, self2, self3}. Possible particle physics examples for bosons constituting scalar DM are weakly interacting slim particles~\cite{ringwald1}, which include axions and axion-like particles~\cite{sikivie1, mielke, sikivie2, yamaguchi, davidson1, berges, davidson2, sikivie3} and spin-1 hidden bosons from string theory~\cite{nelson1, ringwald2}. The Jeans scale associated to the gradient energy of such light scalars is vanishingly small on astrophysical length scales \cite{khlopov}. Below we do not refer to any specific particle physics model, making our subsequent discussion generically applicable to any scalar DM with repulsive self-interaction\footnote{Actually, to ensure the radiative stability of the ultra-light scalar its potential should possess an approximate symmetry which usually gives rise to an attractive self-interaction. However, it is also possible to have a realistic model of an ultra-light scalar with repulsive self-interaction~\cite{self3, berezhiani}.}.

In this paper the distribution of BEC of ultra-light scalar field in outer galactic regions is explored. We argue that scalar DM, together with its success in resolving of the small scale structure $\Lambda$CDM anomalies, can supply the vacuum like equation state inherent to the Dark Energy (DE) governing the contemporary expansion of the universe.

We study distribution around galaxies of a nonlinear complex scalar field $\Phi$ with the Lagrangian,
\begin{equation} \label{L}
L_\Phi = g^{\mu\nu}D_\mu \Phi^* D_\nu \Phi - V(\Phi^*\Phi)~,
\end{equation}
where the self-interacting potential has the form:
\begin{equation} \label{V}
V(\Phi^* \Phi) = \frac {m^2}{2} \Phi^* \Phi + \frac {\lambda}{4} (\Phi^*\Phi)^2 ~.
\end{equation}
The energy momentum tensor of scalar field associate with (\ref{L}) has the usual form:
\begin{eqnarray} \label{T}
T_{\mu\nu} = D_\mu \Phi^* D_\nu \Phi + D_\mu \Phi D_\nu \Phi^* - g_{\mu\nu} L_\Phi  ~.
\end{eqnarray}
Using the Lagrangian (\ref{L}) one can write down the Klein-Gordon equation,
\begin{equation} \label{Klein-Gordon}
\left[\frac{1}{\sqrt{-g}} \partial_\mu \left( \sqrt{-g}g^{\mu\nu}\partial_\nu \right) + m^2 + \lambda N\right] \Phi = 0~.
\end{equation}
The dimensional function,
\begin{equation}\label{N}
N = \Phi^* \Phi ~,
\end{equation}
we call the particle number density, which reduces to the energy and mass densities of the scalar field  under certain conditions \cite{CEMN}.

Since galactic halos seem to be almost spherical, for simplicity we consider the Schwarzschild background,
\begin{equation} \label{Schwarzschild}
ds^2 = f(r) dt^2 - \frac {dr^2}{f(r)} - r^2 \left(d\theta^2 + \sin^2 \theta d\phi^2\right)~,
\end{equation}
where
\begin{equation}
f(r) = 1 - \frac {2M}{r}~.
\end{equation}
In the used system of units, $c = \hbar = 8\pi G =1$, the parameter $2M$ here corresponds to the Schwarzschild horizon and has the dimension of length. We are consider the large scales, $r \gg 2M$, and try to obtain the equations of state corresponding to the solutions of the scalar field equation (\ref{Klein-Gordon}) at different distances from the galactic center. For these regions the back-reactions of the scalar particles into the space-time metric (\ref{Schwarzschild}) can be neglected. In general, to understand the growth of structure and the formation of galaxies the self-gravitating properties of a DM field are essential and more detailed analysis are needed.

The Schwarzschild space-time (\ref{Schwarzschild}) is highly symmetric and to describe galactic haloes we look for spherically symmetric and stationary solutions of the Klein-Gordon equation (\ref{Klein-Gordon}) in the form:
\begin{equation} \label{Phi}
\Phi = e^{i\omega t} \psi (r)~.
\end{equation}
To investigate the cosmological evolution of the system, of course one would need to use a general time-dependent scalar field. The equation (\ref{Klein-Gordon}) for the radial function $\psi (r)$ in (\ref{Phi}) takes the form:
\begin{equation} \label{KG-psi}
\left[f\partial_r^2 + \frac{1+f}{r}\partial_r + \frac {\omega^2}{f} - m^2 - \lambda N \right]\psi(r) = 0~.
\end{equation}
Another simple transformation of the wave function,
\begin{equation} \label{u}
\psi (r) \equiv \frac {u(r)}{r} ~,
\end{equation}
and the introduction of the Regge-Wheeler tortoise coordinate,
\begin{equation}
R = \int \frac {dr}{f} = r + 2M \ln \left( \frac {r}{2M} -1\right)~,
\end{equation}
brings (\ref{KG-psi}) to a non-linear Schr\"{o}dinger-like  equation, which is a kind of Gross-Pitaevskii equation \cite{MGN, FM, DB, CEMN}:
\begin{equation} \label{KG-u}
- u^{''} + \left( V_{eff} + \lambda f N \right) u = \omega^2 u ~,
\end{equation}
where primes denote derivatives with respect to the new radial coordinate $R$ and
\begin{equation} \label{Veff}
V_{eff} = f \left(m^2 + \frac {2M}{r^3}\right) ~.
\end{equation}
Non-linear equations of this type are known for having soliton-like solutions corresponding to BEC, i.e. under some conditions BEC can be obtained from the Klein-Gordon equation~\cite{BH, DKG, BCL, SCB, ML}.

A formal solution of Gross-Pitaevskii equations for the quasi-stationary field distributions can be obtained within so called Thomas-Fermi approximation (by neglecting kinetic energy terms), which is very useful for exploring properties of BEC and is valid for sufficiently large condensate clouds if the system is diluted enough. In Thomas-Fermi approximation, i.e. when $u^{''} \to 0$ in (\ref{KG-u}),
\begin{equation}
\left( V_{eff} + \lambda f N \right) u = \omega^2 u ~,
\end{equation}
and the particle number density function (\ref{N}) obtains the value:
\begin{equation} \label{rhon}
N = \frac {u^2}{r^2} = \frac {\omega^2 - V_{eff} }{\lambda f} ~.
\end{equation}
The size of the scalar field condensate, $d$, which we want to connect with a galactic DM halo, can be calculated from the condition:
\begin{equation} \label{N=0}
N (r = d)= 0 ~.
\end{equation}
For the case of sufficiently large clouds and BEC particles with almost zero momentums,
\begin{equation} \label{d}
d \gg M ~, ~~~~~ R \sim r~, ~~~~~ m \sim \omega~,
\end{equation}
from (\ref{Veff}), (\ref{rhon}) and (\ref{N=0}) we estimate:
\begin{equation} \label{d=}
d \sim \frac 1\omega ~.
\end{equation}
Thus the size of the galactic DM halo (the de Broglie wavelenght of scalar particles) is described by the kinetic term in the Lagrangian (\ref{L}). Demanding coherence up to the galactic scales,
\begin{equation}
d \sim ~{\rm kpc}~,
\end{equation}
from (\ref{d=}) the energy (mass) of the ultra-light scalar particles should be taken around
\begin{equation} \label{m}
\omega \sim m \sim 10^{-24} ~{\rm eV}~,
\end{equation}
which is similar to the bounds used in the papers~\cite{noSelf1, noSelf2, noSelf3, noSelf4, noSelf5}.

For the parameterization of the galactic DM halo energy density distribution, $\rho$, the so called NFW profile is commonly used \cite{NFW-1, NFW-2}. While the parameters and shapes of $\rho$ in different models may differ from the NFW profile~\cite{Moore, BS, PSS, Bertone}, the asymptotic $1/r^3$ behavior in outer galactic regions is widely accepted. The asymptotic value of our effective potential (\ref{Veff}) for the scalar field number density function~(\ref{rhon}) gives
\begin{equation} \label{V-lim}
N_{(r \gg M)} \sim \frac 1\lambda \left(\omega^2 - \frac {2M}{r^3} \right) ~.
\end{equation}
One can see that for large distances the scalar DM density distribution,
\begin{equation} \label{rhoDM}
\rho = T_0^0 = \frac f4\frac {N'^2}{N} + \left( \frac {\omega^2}{f} + m^2\right)N + \frac \lambda 2 N^2 ~,
\end{equation}
which follows from the definition of scalar field energy-momentum tensor (\ref{T}) in the Schwarzschild background space-time (\ref{Schwarzschild}), indeed obeys the NFW behavior.


After this simple estimations of the scalar BEC galactic DM parameter, which appears to be similar with obtained in the papers~\cite{Lee_first,BBM, SRM, shapiro1, shapiro2, shapiro3, MGN, FM, DB, CEMN, BH, DKG, noSelf1, noSelf2, noSelf3, noSelf4, noSelf5}, let us consider the larger scale, $r \gg d$. For these distances galaxies can be imagined as the stationary spherical structures of the DM condensate (\ref{rhoDM}) with some scalar 'tails' in outer galactic region, where the density of DM particle is very low,
\begin{equation} \label{Nto0}
N^2 \to 0 ~,
\end{equation}
while not exactly zero. Then we can neglect the nonlinear term in the Klein-Gordon equation (\ref{KG-psi}), also assume that momentums of scalar particles, together with the angular momentums, are zero and use the dispersion relation:
\begin{equation}\label{omega=m}
\omega^2 = m^2 ~.
\end{equation}
Then in (\ref{KG-psi}) the terms $\omega^2/f$ and $m^2$ cancel each other and the radial Klein-Gordon equation (\ref{KG-psi}) takes the simplest form:
\begin{equation} \label{KG-psi-infty}
\left(\partial_r^2 + \frac{2}{r}\partial_r \right)\psi(r) = 0~.
\end{equation}
Solution to this equation is:
\begin{equation}
\psi_\infty = C ~,
\end{equation}
where $C$ is the integration constant. So, in intergalactic space, we stay with the oscillatory wave function (\ref{Phi}),
\begin{equation} \label{C}
\Phi_\infty \sim C~ e^{\pm imt}~,
\end{equation}
which brings the scalar field Lagrangian (\ref{L}) to the following form:
\begin{equation} \label{L-infty}
L_\infty = - \frac 14 \lambda C^4 ~.
\end{equation}
Therefore, non-zero components of the scalar field energy-momentum tensor (\ref{T}) can be expressed as
\begin{eqnarray} \label{T-DE}
T^0_0 &=& \rho = 2m^2C^2 - L_\infty~, \nonumber \\
T^i_i &=& -p = - L_\infty~.
\end{eqnarray}
In other words, while the size of the galactic DM (\ref{d=}) is described by the kinetic term of order of $\omega^2$, in the outer galactic region dominant is the non-linear part of the DM scalar field potential $\sim L_\infty$. Assuming that
\begin{equation}\label{lambda}
\lambda C^2 \gg 8 m^2 ~,
\end{equation}
one can neglect the first term of $T^0_0$ in (\ref{T-DE}) and use the vacuum-like equation of state:
\begin{equation}
\rho = - p = - L_\infty ~.
\end{equation}
From the expression
\begin{equation}
\rho + 3p = 2C^2\left( m^2 - \frac 14 \lambda C^2\right)
\end{equation}
we see that for the case (\ref{lambda}) the strong energy condition is broken and the universe is in the acceleration regime. Thus, one can identify the asymptotic value of scalar DM potential (\ref{L-infty}) with the cosmological constant $\Lambda$,
\begin{equation} \label{Lamb}
\frac 14 \lambda C^4 \equiv \Lambda \sim 10^{-48}~{\rm GeV^4} ~,
\end{equation}
which is commonly uses as a DE candidate.

This is not surprising, it is well known that in cosmology a slowly-rolling scalar field with a nonzero potential behaves as the DE, since when the time-derivatives of the field can be neglected with respect to the potential the stress-energy tensor in a homogeneous background satisfies the condition $\rho = - p$.

To estimate the values of the integration and coupling constants, $C$ and $\lambda$, note that in outer galactic region the scalar particle number function should be very small (\ref{Nto0}), $N = C^2 \ll 1$. Thus, from (\ref{Lamb}) we find,
\begin{eqnarray} \label{C2}
&10^{-23}~{\rm GeV^2} < C^2 < 0.1~{\rm GeV^2}~, \nonumber \\
&10^{-46}< \lambda < 0.1 ~,
\end{eqnarray}
provided that $\lambda \ll 1$. These values also are compatible with the assumption (\ref{lambda}).

We emphasize that the estimation (\ref{C2}) for the modulus of the scalar DM in the outer galactic region (\ref{C}) is valid only for the Schwarzschild space-time and oscillatory time depended wave function (\ref{Phi}). A similar requirements of the time evolution of the BEC wave function has been applied in~\cite{FM}, while the self-interaction imposed on the constituent ultra-light scalar boson was taken as attractive one. The value of the mass of the ultra light bosons needed for realization of our scenario is of the same order of magnitude as used in the model~\cite{DB}, which also suggest that a macroscopic wavefunction of a BEC gives rise to a vacuum-like state which is typical for a positive cosmological constant. To explore cosmological features of our model one should use the Robertson-Walker metric in intergalactic regions and consider self-gravitating condensates and evolving in time scalar fields. Cosmological studies of the BEC DM show significant differences with respect to the standard cosmology. Presence of such condensates could have modified drastically the cosmological evolution of the early universe, as well as the large scale structure formation process \cite{HM,KL}.


In summary, previously it has been shown in literature that bosons of tiny mass with repulsive self-interaction forming a BEC at early times may account for DM content of the universe. Provided that the BEC (being trapped inside galaxies) can be considered as a system posed into a space-time with Schwarzschild metric, we consider the behavior of its macroscopic wave function characterized by oscillatory time dependence. We observed that for such a setup the strong energy condition is broken providing a vacuum-like equation of state at supra-galactic distances. Thus, ultra-light complex scalar field, which can form large size BEC around galaxies with observed properties of the DM, in principle could supply also the DE component of the universe. These issues need further investigations, since the radius of BEC DM in our simplified model fixed only by a galactic mass, while it is important to describe profiles of galaxies of significantly different sizes. In addition, it is unclear if the DE of the model take all parts of DE in the Universe (e.g. in voids), since it should be Robertson-Walker metric at large scales in the real Universe.

\vskip 3mm

\noindent
{\bf Acknowledgments:} Merab Gogberashvili acknowledges the hospitality of CERN TH Department, where this work has been done. The work of Alexander Sakharov was supported partly by the US National Science Foundation under Grants No.PHY-1402964 and No.PHY-1505463.



\begin{thebibliography}{99}


\bibitem{cdmProb1} B. Moore,
                  {\it "Evidence against dissipationless dark matter from observations of galaxy haloes"},
                  Nature {\bf 370} (1994) 629.

                  DOI: 10.1038/370629a0

\bibitem{cdmProb2} R.A. Flores and J.R. Primack,
                  {\it "Observational and theoretical constraints on singular dark matter halos"},
                  Astrophys. J. {\bf 427} (1994) L1, arXiv: astro-ph/9402004.

                  DOI: 10.1086/187350

\bibitem{coreHalo1} G. Gentile, P. Salucci, U. Klein, D. Vergani and P. Kalberla,
                   {\it "The cored distribution of dark matter in spiral galaxies"},
                   Mon. Not. Roy. Astron. Soc. {\bf 351} (2004) 903, arXiv: astro-ph/0403154.

                   DOI: 10.1111/j.1365-2966.2004.07836.x

\bibitem{cdmProb3} B. Moore, T.R. Quinn, F. Governato, J. Stadel and G. Lake,
                  {\it "Cold collapse and the core catastrophe"},
                  Mon. Not. Roy. Astron. Soc.  {\bf 310} (1999) 1147, arXiv: astro-ph/9903164.

                  DOI: 10.1046/j.1365-8711.1999.03039.x

\bibitem{cdmProb4} M. Boylan-Kolchin, J.S. Bullock and M. Kaplinghat,
                  {\it "Too big to fail? The puzzling darkness of massive Milky Way subhaloes"},
                  Mon. Not. Roy. Astron. Soc. {\bf 415} (2011) L40, arXiv: 1103.0007 [astro-ph.CO].

                  DOI: 10.1111/j.1745-3933.2011.01074.x

\bibitem{cdmProb5} M. Boylan-Kolchin, J.S. Bullock and M. Kaplinghat,
                 {\it "The Milky Way's bright satellites as an apparent failure of LCDM"},
                 Mon. Not. Roy. Astron. Soc. {\bf 422} (2012) 1203, arXiv: 1111.2048 [astro-ph.CO].

                 DOI: 10.1111/j.1365-2966.2012.20695.x

\bibitem{cdmProb6} E. Papastergis, R. Giovanelli, M.P. Haynes and F. Shankar,
                  {\it "Is there a “too big to fail” problem in the field?"},
                  Astron. Astrophys. {\bf 574} (2015) A113, arXiv: 1407.4665 [astro-ph.GA].

                  DOI: 10.1051/0004-6361/201424909

\bibitem{Lee_first} J.-W. Lee and I.-G. Koh,
                  {\it "Galactic halos as boson stars"},
                  Phys. Rev. {\bf D 53} (1996) 2236, arXiv: hep-ph/9507385.

                  DOI: 10.1103/PhysRevD.53.2236

\bibitem{BBM} D. Bertacca, N. Bartolo and S. Matarrese,
             {\it "Unified dark matter scalar field models"},
             Adv. Astron. {\bf 2010} (2010) 904379, arXiv: 1008.0614 [astro-ph.CO].

             DOI: 10.1155/2010/904379

\bibitem{SRM} A. Suarez, V.H. Robles and T. Matos,
             {\it "A Review on the scalar field/Bose-Einstein Condensate dark matter model"},
             Astrophys. Space Sci. Proc. {\bf 38} (2014) 107, arXiv: 1302.0903 [astro-ph.CO].

             DOI: 10.1007/978-3-319-02063-1\_9

\bibitem{shapiro1} B. Li, T. Rindler-Daller and P.R. Shapiro,
                  {\it "Cosmological constraints on Bose-Einstein-Condensed scalar field dark matter"},
                  Phys. Rev. {\bf D 89} (2014) 083536, arXiv: 1310.6061 [astro-ph.CO].

                  DOI: 10.1103/PhysRevD.89.083536

\bibitem{shapiro2} B. Li, P.R. Shapiro and T. Rindler-Daller,
                   {\it "Bose-Einstein-condensed scalar field dark matter and the gravitational wave
                   background from inflation: new cosmological constraints and its detectability by LIGO"},
                   Phys. Rev. {\bf D 96} (2017) 063505, arXiv: 1611.07961 [astro-ph.CO].

	               DOI: 10.1103/PhysRevD.96.063505

\bibitem{shapiro3} T. Rindler-Daller and P.R. Shapiro,
                  {\it "Complex scalar field dark matter on galactic scales"},
                  Mod. Phys. Lett. {\bf A 29} (2014) 1430002, arXiv: 1312.1734 [astro-ph.CO].

                  DOI: 10.1142/S021773231430002X

\bibitem{MGN} T. Matos, F.S. Guzman and D. Nunez,
             {\it "Spherical scalar field halo in galaxies"},
             Phys. Rev. {\bf D 62} (2000) 061301, arXiv: astro-ph/0003398.

             DOI: 10.1103/PhysRevD.62.061301

\bibitem{FM} T. Fukuyama and M. Morikawa,
            {\it "Relativistic Gross-Pitaevskii equation and the cosmological Bose Einstein condensation:
            Quantum structure in universe"},
            J. Phys. Conf. Ser. {\bf 31} (2006) 139, arXiv: astro-ph/0509789.

            DOI: 10.1088/1742-6596/31/1/023

\bibitem{DB} S. Das and R.K. Bhaduri,
            {\it "Dark matter and dark energy from a Bose-Einstein condensate"},
            Class. Quant. Grav. {\bf 32} (2015) 105003, arXiv: 1411.0753 [gr-qc].

            DOI: 10.1088/0264-9381/32/10/105003

\bibitem{CEMN} E. Castellanos, C. Escamilla-Rivera, A. Macias and D. Nunez,
               {\it "Scalar field as a Bose-Einstein Condensate?"},
               JCAP {\bf 1411} (2014) 034, arXiv: 1310.3319 [gr-qc].

               DOI: 10.1088/1475-7516/2014/11/034

\bibitem{BH} C.G. Boehmer and T. Harko,
            {\it "Can dark matter be a Bose-Einstein condensate?"},
            JCAP {\bf 0706} (2007) 025, arXiv: 0705.4158 [astro-ph].

            DOI: 10.1088/1475-7516/2007/06/025

\bibitem{DKG} M. Dwornik, Z. Keresztes and L.A. Gergely,
             {\it "Bose-Einstein Condensate Dark Matter Halos confronted with galactic rotation curves"},
             Adv. High Energy Phys. {\bf 2017} (2017) 4025386, arXiv: 1406.0388 [astro-ph.GA].

             DOI:  	10.1155/2017/4025386

\bibitem{noSelf1} S.J. Sin,
                 {\it "Late time cosmological phase transition and galactic halo as Bose liquid"},
                 Phys. Rev. {\bf D 50} (1994) 3650, arXiv: hep-ph/9205208.

                 DOI: 10.1103/PhysRevD.50.3650

\bibitem{noSelf2} W. Hu, R. Barkana and A. Gruzinov,
                 {\it "Cold and fuzzy dark matter"},
                 Phys. Rev. Lett. {\bf 85} (2000) 1158, arXiv: astro-ph/0003365.

                 DOI: 10.1103/PhysRevLett.85.1158

\bibitem{noSelf3} T. Matos, A. Vazquez-Gonzalez and J. Magana,
                 {\it "$\phi^2$ as dark matter"},
                 Mon. Not. Roy. Astron. Soc.  {\bf 393} (2009) 1359, arXiv: 0806.0683 [astro-ph].

                 DOI: 10.1111/j.1365-2966.2008.13957.x

\bibitem{noSelf4} J.W. Lee and S. Lim,
                 {\it "Minimum mass of galaxies from BEC or scalar field dark matter"},
                 JCAP {\bf 1001} (2010) 007, arXiv: 0812.1342 [astro-ph].

                 DOI: 10.1088/1475-7516/2010/01/007

\bibitem{noSelf5} V. Lora, J. Magana, A. Bernal, F.J. Sanchez-Salcedo and E.K. Grebel,
                 {\it "On the mass of ultra-light bosonic dark matter from galactic dynamics"},
                 JCAP {\bf 1202} (2012) 011, arXiv: 1110.2684 [astro-ph.GA].

                 DOI: 10.1088/1475-7516/2012/02/011

\bibitem{self1} P.J.E. Peebles,
               {\it "Fluid dark matter"},
               Astrophys. J. {\bf 534} (2000) L127, arXiv: astro-ph/0002495.

               DOI: 10.1086/312677

\bibitem{Goodman1} J. Goodman,
                  {\it "Repulsive dark matter"},
                  New Astron. {\bf 5} (2000) 103, arXiv: astro-ph/0003018.

                  DOI: 10.1016/S1384-1076(00)00015-4

\bibitem{Julian1} A. Arbey, J. Lesgourgues and P. Salati,
                 {\it "Galactic halos of fluid dark matter"},
                 Phys. Rev. {\bf D 68} (2003) 023511, arXiv: astro-ph/0301533.

                 DOI: 10.1103/PhysRevD.68.023511

\bibitem{self2} J. Eby, C. Kouvaris, N.G. Nielsen and L.C.R. Wijewardhana,
               {\it "Boson stars from self-interacting dark matter"},
               JHEP {\bf 1602} (2016) 028,  arXiv: 1511.04474 [hep-ph].

               DOI: 10.1007/JHEP02(2016)028

\bibitem{self3} J. Fan,
               {\it "Ultralight repulsive dark matter and BEC"},
               Phys. Dark Univ. {\bf 14} (2016) 84, arXiv: 1603.06580 [hep-ph].

               DOI: 10.1016/j.dark.2016.10.005

\bibitem{ringwald1} A. Ringwald,
                   {\it "Exploring the role of axions and other WISPs in the dark universe"},
                   Phys. Dark Univ. {\bf 1} (2012) 116, arXiv: 1210.5081 [hep-ph].

                   DOI: 10.1016/j.dark.2012.10.008

\bibitem{sikivie1} P. Sikivie and Q. Yang,
                  {\it "Bose-Einstein condensation of dark matter axions"},
                  Phys. Rev. Lett. {\bf 103} (2009) 111301, arXiv: 0901.1106 [hep-ph].

                  DOI: 10.1103/PhysRevLett.103.111301

\bibitem{mielke} E.W. Mielke and J.A.V. Perez,
                {\it "Axion condensate as a model for dark matter halos"},
                Phys. Lett. {\bf B 671} (2009) 174.

                DOI: 10.1016/j.physletb.2008.11.044

\bibitem{sikivie2} O. Erken, P. Sikivie, H. Tam and Q. Yang,
                 {\it "Cosmic axion thermalization"},
                 Phys. Rev. {\bf D 85} (2012) 063520, arXiv: 1111.1157 [astro-ph.CO].

                 DOI: 10.1103/PhysRevD.85.063520

\bibitem{yamaguchi} K. Saikawa and M. Yamaguchi,
                   {\it "Evolution and thermalization of dark matter axions in the condensed regime"},
                   Phys. Rev. {\bf D 87} (2013) 085010, arXiv: 1210.7080 [hep-ph].

                   DOI: 10.1103/PhysRevD.87.085010

\bibitem{davidson1} S. Davidson and M. Elmer,
                    {\it "Bose Einstein condensation of the classical axion field in cosmology?"},
                    JCAP {\bf 1312} (2013) 034, arXiv: 1307.8024 [hep-ph].

                    DOI: 10.1088/1475-7516/2013/12/034

\bibitem{berges} J. Berges and J. Jaeckel,
                {\it "Far from equilibrium dynamics of Bose-Einstein condensation for axion dark matter"},
                Phys. Rev. {\bf D 91} (2015) 025020, arXiv: 1402.4776 [hep-ph].

                DOI: 10.1103/PhysRevD.91.025020

\bibitem{davidson2} S. Davidson,
                   {\it "Axions: Bose Einstein Condensate or classical field?"},
                   Astropart. Phys. {\bf 65} (2015) 101, arXiv: 1405.1139 [hep-ph].

                   DOI: 10.1016/j.astropartphys.2014.12.007

\bibitem{sikivie3} N. Banik, A.J. Christopherson, P. Sikivie and E.M. Todarello,
                  {\it "New astrophysical bounds on ultralight axionlike particles"},
                  Phys. Rev. {\bf D 95} (2017) 043542, arXiv: 1701.04573 [astro-ph.CO].

                  DOI: 10.1103/PhysRevD.95.043542

\bibitem{nelson1} A.E. Nelson and J. Scholtz,
                 {\it "Dark light, dark matter and the misalignment mechanism"},
                 Phys. Rev. {\bf D 84} (2011) 103501, arXiv: 1105.2812 [hep-ph].

                 DOI: 10.1103/PhysRevD.84.103501

\bibitem{ringwald2} P. Arias, D. Cadamuro, M. Goodsell, J. Jaeckel, J. Redondo and A. Ringwald,
                   {\it "WISPy cold dark matter"},
                   JCAP {\bf 1206} (2012) 013, arXiv: 1201.5902 [hep-ph].

                   DOI: 10.1088/1475-7516/2012/06/013

\bibitem{khlopov} M. Khlopov, B.A. Malomed and I.B. Zeldovich,
                 {\it "Gravitational instability of scalar fields and formation of primordial black holes"},
                 Mon. Not. Roy. Astron. Soc. {\bf 215} (1985) 575.

                 DOI: 10.1093/mnras/215.4.575

\bibitem{berezhiani} L. Berezhiani and J. Khoury,
                    {\it "Theory of dark matter superfluidity"},
                    Phys. Rev. {\bf D 92} (2015) 103510, arXiv: 1507.01019 [astro-ph.CO].

                    DOI: 10.1103/PhysRevD.92.103510


\bibitem{BCL} D. Bettoni, M. Colombo and S. Liberati,
             {\it "Dark matter as a Bose-Einstein Condensate: the relativistic non-minimally coupled case"},
             JCAP {\bf 02} (2014) 004, arXiv: 1310.3753 [astro-ph.CO].

             DOI: 10.1088/1475-7516/2014/02/004

\bibitem{SCB} H. Schive, T. Chiueh and T. Broadhurst,
              {\it "Cosmic structure as the quantum interference of a coherent dark wave"},
              Nature Phys. {\bf 10} (2014) 496, arXiv: 1406.6586 [astro-ph.GA].

              DOI: 10.1038/nphys2996

\bibitem{ML} T. Matos and L.A. Urena-Lopez,
            {\it "A further analysis of a cosmological model of quintessence and scalar dark matter"},
            Phys. Rev. {\bf D 63} (2001) 063506, arXiv: astro-ph/0006024.

            DOI: 10.1103/PhysRevD.63.063506


\bibitem{NFW-1} J.F. Navarro, C.S. Frenk and S.D.M. White,
               {\it "The structure of cold dark matter halos"},
               Astrophys. J. {\bf 462} (1996) 563, arXiv: astro-ph/9508025;

               DOI: 10.1086/177173

\bibitem{NFW-2} J.F. Navarro, C.S. Frenk and S.D.M. White,
               {\it "A universal density profile from hierarchical clustering"},
               Astrophys. J. {\bf 490} (1997) 493, arXiv: astro-ph/9611107.

               DOI: 10.1086/304888

\bibitem{Moore} B. Moore, F. Governato, T. Quinn, J. Stadel and G. Lake,
               {\it "Resolving the structure of cold dark matter halos"},
               Astrophys. J. {\bf 499} (1998) L5, arXiv: astro-ph/9709051.

               DOI: 10.1086/311333

\bibitem{BS} J.N. Bahcall and R.M. Soneira,
            {\it "The universe at faint magnitudes. I - Models for the galaxy and the predicted star counts"},
            Astrophys. J. Suppl. Series {\bf 44} (1980) 73.

            DOI: 10.1086/190685

\bibitem{PSS} M. Persic, P. Salucci and F. Stel,
             {\it "The universal rotation curve of spiral galaxies: 1. The Dark matter connection"},
             Mon. Not. Roy. Astron. Soc. {\bf 281} (1996) 27, arXiv: astro-ph/9506004.

             DOI: 10.1093/mnras/281.1.27

\bibitem{Bertone} G. Bertone,
                 {\it Particle Dark Matter: Observations, Models and Searches}
                 (Cambridge University Press, 2010).


\bibitem{HM} T. Harko and G. Mocanu,
            {\it "Cosmological evolution of finite temperature Bose-Einstein Condensate dark matter"},
            Phys. Rev. {\bf D 85} (2012) 084012, arXiv: 1203.2984 [gr-qc].

            DOI: 10.1103/PhysRevD.85.084012

\bibitem{KL} B. Kain and H.Y. Ling,
            {\it "Cosmological inhomogeneities with Bose-Einstein Condensate dark matter"},
            Phys. Rev. {\bf D 85} (2012) 023527, arXiv: 1112.4169 [hep-ph].

            DOI: 10.1103/PhysRevD.85.023527

\end{thebibliography}
\end{document}